\documentclass[aps,prl,reprint,groupedaddress]{revtex4-1}
\usepackage{epsf}
\usepackage{epsfig}
\usepackage{graphicx}
\usepackage{color}
\usepackage{siunitx}
\usepackage{epstopdf}
\begin{document}
\title {\large
Sub-Cycle Interference upon Tunnel-Ionization by Counterrotating Two-Color Fields}
\author{S. Eckart$^{1}$}
\email{eckart@atom.uni-frankfurt.de}
\author{M. Kunitski$^{1}$}
\author{I. Ivanov$^{2}$}
\author{M. Richter$^{1}$}
\author{K. Fehre$^{1}$}
\author{A. Hartung$^{1}$}
\author{J. Rist$^{1}$}
\author{K. Henrichs$^{1}$}
\author{D. Trabert$^{1}$}
\author{N. Schlott$^{1}$}
\author{L. Ph. H. Schmidt$^{1}$}
\author{T. Jahnke$^{1}$}
\author{M. S. Sch\"offler$^{1}$}
\author{A. Kheifets$^{1,2}$}
\author{R.~D\"{o}rner$^{1}$}
\email{doerner@atom.uni-frankfurt.de}
\affiliation{$^1$ Institut f\"ur Kernphysik, Goethe-Universit\"at, Max-von-Laue-Str. 1, 60438 Frankfurt, Germany \\
$^2$ Research School of Physics and Engineering, The Australian National University, Canberra ACT 0200, Australia.}
\date{\today}
\begin{abstract}
We report on three-dimensional (3D) electron momentum distributions from single ionization of helium by a laser pulse consisting of two counterrotating circularly polarized fields ($\SI{390}{\nano\meter}$ and $\SI{780}{\nano\meter}$). A pronounced 3D low energy structure and sub-cycle interferences are observed experimentally and reproduced numerically using a trajectory based semi-classical simulation. The orientation of the low energy structure in the polarization plane is verified by numerical simulations solving the time dependent Schr\"odinger equation.
\end{abstract}
\maketitle
An electron wave packet is born when a strong laser field ionizes an atom. This wave packet is then driven by the strong optical field. Since it is possible to control and shape optical waveforms extremely well they can be used to steer the electronic wave packets in the presence of the Coulomb field of the residual ion. This makes the process of strong field ionization an exquisite playground for the exploration of wave packet quantum dynamics. The process becomes particularly transparent in the case of optical tunnel ionization \cite{Keldysh1965} where the propagation of the electron after tunneling can be successfully modeled classically. This allows to interpret structures in the observed electron momentum space distribution as interference of different classical paths that lead to the same final electron momentum \cite{Paulus2005,Richter2015,Zhang2014}. If the ionization times are separated by the oscillatory period of the ionizing laser pulse, inter-cycle interferences appear as peaks in the electron energy spectrum that are separated by the photon energy. This process is known as above threshold ionization. A second class of structures arises from interferences between wave packets or parts of wave packets which are launched within the same cycle of the driving field. These are known as sub-cycle or intra-cycle interferences \cite{Paulus2005, Arbo2010} and have been proposed to probe the electron dynamics and correlations, as well as atomic and molecular potentials on ultrafast time scales \cite{Gopal2009}. For circularly polarized light sub-cycle interferences cannot occur because the angle in the plane of polarization unambiguously encodes the laser phase at ejection of the electron \cite{Eckle2008,Dietrich_2000}. For linearly polarized light sub-cycle interference is affected by the Coulomb attraction of the parent ion \cite{Huismans2011, Zhang2014} or molecule \cite{Haertelt2016,Meckel2014}  since different paths of the electron wave packets are affected differently by the ionic potential \cite{Meckel2008}. If a laser pulse is composed of two harmonic colors many further control parameters emerge. One class of such combined electric fields are orthogonal two-color fields that already support sub-cycle interferences since there are two birth-times with the same vector potential \cite{Geng2015,Richter2015,Zhang2014,He2016} and even allow to retrieve properties of the valence-electron cloud in atoms \cite{Xie2012}.

Extremely well controllable waveforms of non-trivial shape are generated by counterrotating circularly polarized two-color fields (CRTC) \cite{Mancuso2015}. Originally such waveforms have been proposed to tailor the polarization of high order harmonics \cite{becker1999schemes,fleischer2014spin} since in CRTC fields electrons with high kinetic energy can recollide with their parent ions \cite{mancuso2016controlling}. Previous experiments investigated the energy and the sub-cycle timing of such recolliding electrons \cite{Eckart2016, chaloupka2016, Mancuso2016PRL}. In those works also the influence of the Coulomb potential and its role regarding the creation of low energy electrons - which are absent for ionization by circularly polarized laser pulses - has been observed \cite{Coulomb_focusing}. 

In the present Letter we use these waveforms and investigate 3D electron momentum distributions from single ionization of helium with unprecedented resolution and statistical significance. This allows us to discover novel structures unseen in previous works. We show that these structures are due to sub-cycle interferences between two classes of electron wave packets which are born having unequal vector potentials and interfere only because of the interplay of their initial momenta and Coulomb interaction with the parent ion. Furthermore, we explain the origin of the low energy structure and how it can be used to experimentally determine the orientation of the electric field in the laboratory frame which is an important insight for future two-color attoclock \cite{Eckle2008,Beaulieu} experiments.

\begin{figure}
\epsfig{file=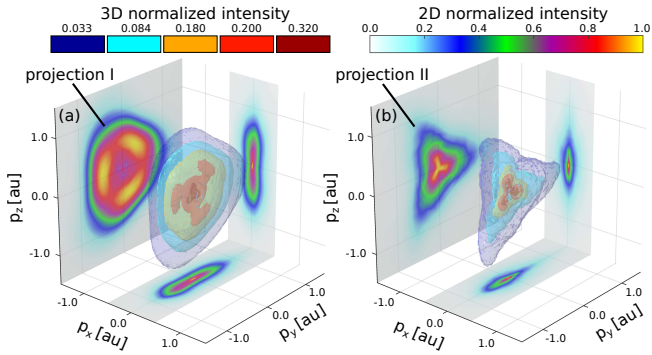, width=8.7cm} 
\caption{Three-dimensional (3D) electron momentum distributions from ionization of helium by counterrotating two-color fields are shown. The 3D momentum distributions are visualized by combining five semitransparent isosurfaces encoding the intensity of the 3D histogram by color. The $p_yp_z$-plane is the polarization plane. The 2D histograms show projections. a) The field ratio of the two colors and their combined intensity are $E_{390}/E_{780}=0.84$ and $I=\SI{8.14E+14}{\watt\per\centi\meter^2}$. Note the pronounced (red) inner low energy structure in the 3D distribution which is almost invisible in projection I. b) shows the same as (a) but for $E_{390}/E_{780}=1.20$ and $I=\SI{6.75E+14}{\watt\per\centi\meter^2}$.}
\label{fig_figure1} 
\end{figure}

The field parameters we use and the resulting 3D electron momentum distributions are shown in Fig. \ref{fig_figure1}. The three-fold symmetry of the field is imparted on the momentum distribution. Comparing the 3D structure with its projection onto the plane of polarization shows that even though the field is only two-dimensional (2D), the third dimension of the electron momentum distribution is highly non trivial. Much of the rich 3D structure is lost by integrating out this third dimension as can be seen most graphically in panel a) of Fig. \ref{fig_figure1}. Here a high density in the inner region (low energy structure) is observable which is hardly visible in projection I. In contrast, Fig. \ref{fig_figure1} (b) exhibits more electrons with small momenta that are visible in 3D as well as in projection II. Thus, full understanding of the wave packet dynamics requires a complete analysis of 3D data. 

In order to generate two-color fields we used a $200$ $\mu\text{m}$ BBO to frequency double a $\SI{780}{\nano\meter}$ laser pulse (KMLabs Dragon, $\SI{40}{\femto\second}$ FWHM, $\SI{8}{\kilo\hertz}$). The fundamental and the second harmonic were separated with a dielectric beam splitter. Before the two were recombined, a neutral density filter, followed by a lambda-quarter and a lambda-half waveplate, was installed in each pathway. A nm-delay stage in the arm of the fundamental wavelength was used to adjust the temporal overlap (i.e. relative phase) between the two colors \cite{Richter2015}. A spherical mirror ($f=\SI{80}{\milli\meter}$) focused the laser field (aperture of $\SI{8}{\milli\meter}$ ($\SI{5}{\milli\meter}$) for $\SI{780}{\nano\meter}$ ($\SI{390}{\nano\meter}$)) into a helium target that was generated using supersonic gas expansion. The intensity in the focus was determined for both colors separately analyzing the photoelectron momentum distributions from ionization by circularly polarized light as in \citep{Eckart2016}. We estimate the uncertainty of the absolute intensity for $\SI{780}{\nano\meter}$ and $\SI{390}{\nano\meter}$ to be $10\,\%$ and $20\,\%$, respectively. Since our intensity calibration is an \textit{in situ} measurement of the electric field, it inherently includes the focal averaging for single ionization (Rayleigh length of about $\SI{150}{\micro\meter}$, jet diameter $\SI{1}{\milli\meter}$). The 3D electron momentum distributions from single ionization of helium presented in this paper have been measured in coincidence with helium ions using cold-target recoil-ion momentum spectroscopy (COLTRIMS) \cite{ullrich2003recoil}. The length of the electron and ion arm was $\SI{378}{\milli\meter}$ and $\SI{67.8}{\milli\meter}$ respectively. Homogeneous electric and magnetic fields of $\SI{10.72}{\volt\per\centi\meter}$ and $8.4\,$G, respectively, guided electrons and ions towards position sensitive microchannel plate detectors with three layer delay-line anodes \cite{jagutzki2002multiple}. The experimental setup was the same as in \cite{Eckart2016}.

\begin{figure}
\epsfig{file=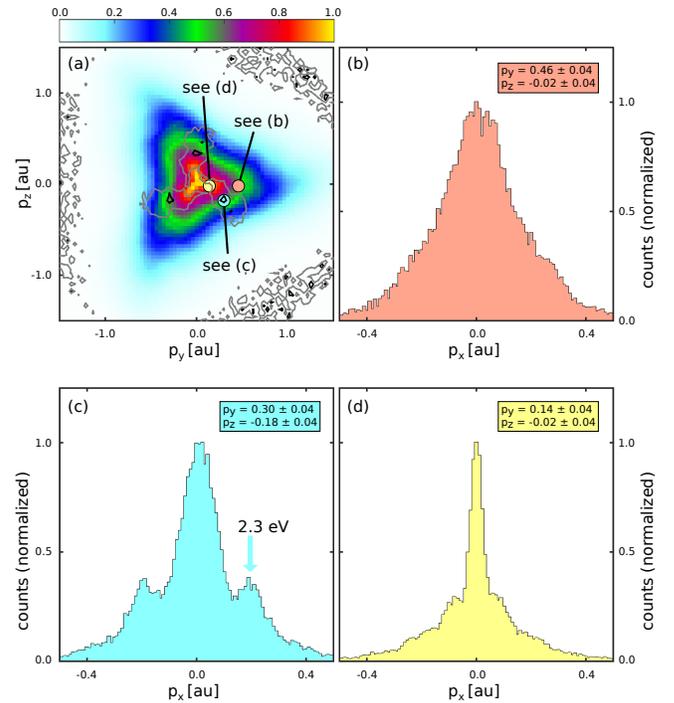, width=8.7cm} 
\caption{The momentum distribution in the light propagation direction is investigated for various selected momenta in the plane of polarization. (a) shows  projection II from Fig. \ref{fig_figure1}. The dots in the polarization plane in (a) indicate the used condition for panel (b), (c), (d) that show the momentum ($p_x$) in the light propagation direction. (b) shows a Gaussian-like distribution (as for circularly polarized light), (c) shows sub-cycle interference (fringes) and (d) shows the cusp-like initial momentum distribution (as for linear light). The side-peaks of the sub-cycle interference in (c) correspond to an electron energy of $\SI{2.3}{\electronvolt}$. The fringe visibility is indicated by contour lines in (a). Gray contour lines indicate 50 \% fringe visibility compared to the black contour line.}
\label{fig_figure2} 
\end{figure}

Fig. \ref{fig_figure2} visualizes the same data set as shown in Fig. \ref{fig_figure1} (b), but in more detail. Figs. \ref{fig_figure2} (b), (c), (d) depict the momentum distribution along the light propagation direction ($p_x$, transverse electron momentum) for well defined values of $p_y$ and $p_z$ (i.e. the momentum density in a narrow column along the light propagation direction). Near the tip of the three-fold structure in the plane of polarization a Gaussian-like momentum distribution along $p_x$ is found (b). Such distributions are known for circularly polarized light \cite{Arissian2010, Ivanov2016} where the interaction with the ionic core upon ionization is minimal. Fig. \ref{fig_figure2} (d) shows the distribution for selected low momenta on the yellow three armed star which has a pronounced cusp at zero momentum. The same evolution of the transverse electron momentum distributions  from Gaussian  to cusp is reproduced by the TDSE calculations as in \cite{Ivanov2016}. Similar cusp structures are known to be caused by Coulomb focusing in the case of ionization by linearly polarized light \cite{Coulomb_focusing,Rudenko2005}. 

\begin{figure}
\epsfig{file=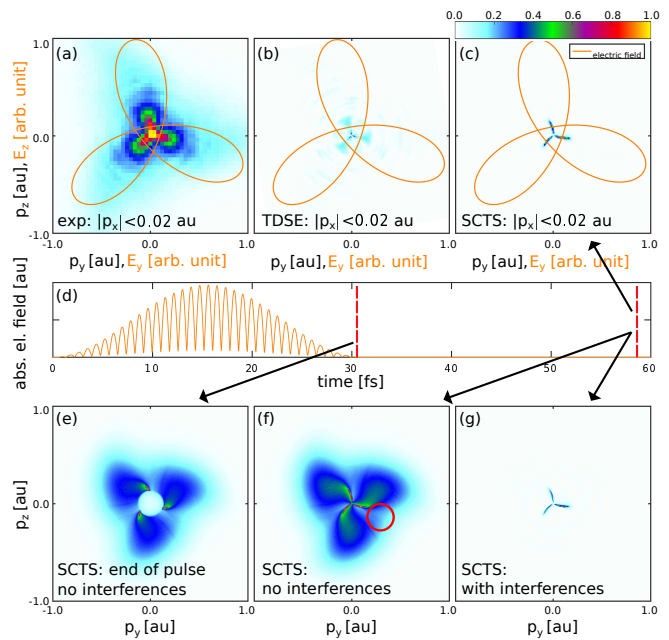, width=8.7cm} 
\caption{Comparison of the experiment (a), a solution of the TDSE (b) and the SCTS result (c) for small momenta in the light propagation direction $|p_x|<0.02 \,$a.u. shows a very good agreement. The temporal evolution of the laser field magnitude used in the SCTS model is shown in (d). (e) shows the electron momentum distribution from the SCTS immediately after the end of the laser pulse (without interferences and only electrons with positive total energy are shown). In (f) the same electrons are presented for asymptotically large time. It can be graphically seen that the Coulomb potential slows down the electrons after the laser pulse is over. This is the reason for the pronounced low energy structure in CRTC fields. (g) is the same as (f) but includes interferences. The red circle in (f) guides the eye to the overlapping region of two neighboring lobes. The momentum component which is not shown ($p_x$) is projected out in (e), (f), (g).}
\label{fig_figure3}
\end{figure}

In order to unravel the origin of the different distribution shapes observed in Figs. \ref{fig_figure2} (b), (d) we have performed a semi-classical two step (SCTS) simulation following the procedure described in  \cite{Shilovski2016} using 500 million trajectories. The classical electron trajectories were calculated with weights obtained by the ADK theory \cite{Delone1991} starting at the exit of the tunnel with zero momentum in the direction parallel to the tunnel-direction and a Gaussian momentum distribution transverse to it. The electron trajectories are calculated in the presence of the Coulomb field and the semi-classical phase accumulates during propagation. The low energy structure within a thin slice ($|p_x|<0.02 \,$a.u.) in the light propagation direction from experiment and theory is compared in Fig. \ref{fig_figure3} (a), (b), (c). The result obtained semi-classically is shown in Fig. \ref{fig_figure3} (c) and its orientation agrees well with the result from solution of the time dependent Schr\"odinger equation (TDSE, Fig. \ref{fig_figure3} (b)).  The laser parameters for the TDSE and the SCTS simulations are the same as in Fig. \ref{fig_figure1} (b) and the magnitude of the laser field used for the SCTS simulation is shown in Fig. \ref{fig_figure3} (d).

Since time evolution along each trajectory is known in the SCTS model, the buildup of the low energy structure can be followed in time. Fig. \ref{fig_figure3} (e) shows the simulated 2D electron momentum distribution in the plane of polarization just after the end of the laser pulse. Only electrons with positive total energy are shown explaining the void at low energy. The electron momentum distribution in the asymptotic limit of ${t\to\infty}$ is presented in Fig. \ref{fig_figure3} (f) which is determined by analytic Coulomb mapping taking advantage of energy conservation after the end of the laser pulse. The origin of the low energy structure in the final momentum distribution is evidently due to the strong Coulomb attraction, which sucks in slow electrons towards zero momentum. This attraction is very significant for small momenta explaining the cusp in Fig. \ref{fig_figure2} (d) and the angular offset between the lobes of the propeller field and the angles that the low energy structure points to in Fig. \ref{fig_figure2} (a). In Fig. \ref{fig_figure3} (e), (f) the purely classical results are shown. Fig. \ref{fig_figure3} (g) shows the same calculation but taking the semi-classical phase into account 
\cite{Shilovski2016}.

The excellent agreement of the orientation of the three armed star relative to the field orientation in the classical and TDSE calculation can now be used to  obtain the experimental field orientation. Using the SCTS model we have also tested the orientation of the three armed star against unrealistically high variations of intensity (up to +20\,\% for both intensities), field ratio of the two colors (changed by a factor of up to $1.2$), ellipticity of the two individual colors (we have tested $\epsilon=0.8$ for the second harmonic and $\epsilon=0.9$ for the fundamental with the main axis of the two ellipses beeing orthogonal). We have changed the initial momenta in direction of the tunnel (up to $0.3$\,a.u.). In all those cases we have found that the absolute orientation of the three armed star is robust to better than $\SI{2}{\degree}$ against these variations. The robustness of this prominent feature can therefore be used to infer the lab frame orientation of the experimental propeller field in the focus.

\begin{figure}
\epsfig{file=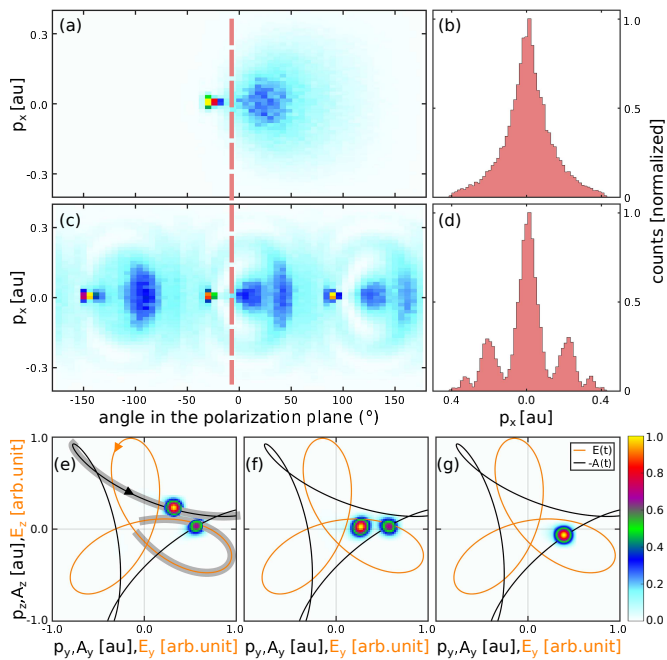, width=8.7cm} 
\caption{The origin of the sub-cycle interferences is investigated using the semi-classical SCTS simulation. Only electrons with an energy of $2.3\pm\SI{0.1}{\electronvolt}$ are selected. The resulting subset is presented as 2D momentum distributions in (a), (c) for ionization within one lobe of the combined electric field (indicated as gray shaded phases of the electric field and the vector potential in (e)) and a full single cycle of the combined electric field respectively. Panels (b), (d) show subsets from (a), (b) selecting the angle in the polarization plane to be $-6 \pm \SI{6}{\degree}$. Manifestly panel (d) shows modulations which panel (b) does not show. This indicates that the observed modulations are indeed sub-cycle interferences. Tracing back all electrons by showing their negative vector potential (panel (e)) reveals their corresponding ``birth times''. Those times/phases are seen as dots on the vector potential in panel (e). Panel (f) shows the sum of the negative vector potential at the instant of ionization and the initial momentum of the electrons just after tunneling. Panel (g) shows the final momentum. The data shown in (e), (f), (g) has been convoluted with a Gaussian distribution to facilitate visibility. The arrows in (e) indicate the time evolution of the electric field and the negative vector potential.}
\label{fig_figure4}
\end{figure}

We now return to Fig. \ref{fig_figure2}. Surprisingly, a pronounced oscillatory modulation on top of the broad peak is found in the intermediate region between two neighboring lobes in the plane of polarization (see Fig. \ref{fig_figure2} (c)). On closer examination a visible reminiscence of an oscillatory structure is evident in Fig. \ref{fig_figure2} (d) as well. Such sinusoidal oscillations of probability density are clear indications of two path interferences. This is a  surprising result because, unlike in the case of linearly polarized light \cite{Paulus2005}, it is not possible to find two contributing ``birth times'' that have the same vector potential within one cycle of the used two-color field. 

To obtain a measure for the fringe-visibility for each combination of $p_y$ and $p_z$ the sum of the momentum distribution in $p_x$-direction is normalized to one and a Gaussian fit is subtracted.  The remainder is Fourier transformed. The fringe-visibility is defined as the product of the amplitude of the Fourier component corresponding to 4 oscillations per a.u. and the integral of the remainder. Contour lines in Fig. \ref{fig_figure2} (a) visualize the fringe-visibility for the data from Fig. \ref{fig_figure1} (b). For the other set of laser parameters (as in Fig. \ref{fig_figure1} (a)) and momenta above (below) 0.3 a.u. in the plane of polarization the fringe-visibility is below 10\% (50\% ) of the value of the maximum fringe-visibility of Fig. \ref{fig_figure2} (a).

To reveal the origin of those fringes we again refer to the SCTS simulation. The phase space region showing the oscillations in Fig. \ref{fig_figure2} (c) corresponds to electrons with an energy of $\SI{2.3}{\electronvolt}$. We thus postselect electrons with electron energies in the range $\SI{2.2}{\electronvolt}$ to $\SI{2.4}{\electronvolt}$ in our SCTS simulation. If we in addition restrict our postselection to trajectories ``born'' within only one single lobe of the three lobe propeller field as shown in Fig. \ref{fig_figure4} (a), (b) no oscillatory structure arises, suggesting that it results from interferences between trajectories originated from two neighboring lobes of the propeller. This is readily confirmed in Fig. \ref{fig_figure4} (c), (d), showing a calculation where we have restricted the ionization time to one entire cycle of the $\SI{780}{\nano\meter}$ field. This proves that those structures result from sub-cycle but inter-propeller-lobe interferences. In Fig. \ref{fig_figure4} (e), (f), (g) the origin of this interference is investigated using the trajectory information of our semi-classical simulation. In this calculation the final momentum is the sum of the negative vector potential at the instant of ionization, the initial momentum of the electron upon ionization and the momentum changes induced by the Coulomb interaction of the liberated electron with its parent ion. Only those trajectories can interfere where these three factors conspire to yield the identical final momentum. In the series Figs. \ref{fig_figure4} (e), (f), (g) the individual contributions of these three factors is traced. It shows that the interfering trajectories originate from tunneling at phases where the vector potential is rather different and it is only the initial momentum and the Coulomb interaction which finally make these trajectories lead to the same final momentum.

In conclusion, we have shown that the electron momentum component along the light propagation direction upon ionization by CRTC fields shows a rich structure. Depending on the momentum component in the plane of polarization either a cusp-like distribution (known for linearly polarized light), a Gaussian distribution (known for circularly polarized light) or sub-cycle interferences are observed. These interferences are robust with respect to volume averaging. In our semi-classical picture the overlap in final momentum of the contributing wave packets can be explained only in a model that includes the electrons's Coulomb interaction with the ionic core and its initial momentum distribution at the tunnel exit. Sub-cycle interference could be used to probe ultrafast processes in atoms and molecules while CRTC fields allow to observe such interferences in an otherwise empty region in momentum space, allowing for high modulation depths as compared to linear light \cite{Gopal2009} without the need for carrier-envelope phase stable pulses. Further, the origin of the low energy structure is investigated and it is shown that is can be used to precisely determine the orientation of the combined electric field which will be useful in future attoclock experiments employing CRTC fields. One particular promising application is the use of CRTC fields and effects described in the present paper for exploring chiral molecules. Such experiments are currently underway \cite{Beaulieu}.

\section{Acknowledgments}
\normalsize
This work was supported by the DFG Priority Programme ``Quantum Dynamics in Tailored Intense Fields''. A.H. and K.H. acknowledge support by the German merit foundation. A.K. acknowledges support by the Wilhelm and Else Heraeus Foundation.

\bibliographystyle{apsrev4-1}
\end{document}